\documentclass{emulateapj} 
\begin{document} 
\title{On Estimating the Mass of Keplerian Accretion Disks in H$_{2}$O Maser
Galaxies}                                                               
 
 \author{C. Y. Kuo\altaffilmark{1}, M. J. Reid\altaffilmark{2}, J.
A. Braatz\altaffilmark{3}, F. Gao\altaffilmark{3}, C. M. V. Impellizzeri\altaffilmark{3,4},
W. T. Chien\altaffilmark{1}}                                            
 
\affil{\altaffilmark{1}Physics Department, National Sun Yat-Sen University,
No. 70, Lien-Hai Rd, Kaosiung City 80424, Taiwan, R.O.C }               
\affil{\altaffilmark{2}Harvard-Smithsonian Center for Astrophysics, 
60 Garden Street, Cambridge, MA 02138, USA} 
\affil{\altaffilmark{3}National Radio Astronomy Observatory, 520 
Edgemont Road, Charlottesville, VA 22903, USA} 
\affil{\altaffilmark{4}Joint ALMA Office, Alonso de Córdova 3107,
Vitacura, Santiago, Chile}

\begin{abstract} 
H$_{2}$O maser disks with Keplerian rotation in active galactic nuclei offer a clean way to determine accurate black hole mass and the Hubble constant. An important assumption made in using a Keplerian H$_{2}$O maser disk for measuring the black hole mass and the Hubble constant is that the disk mass is negligible compared to the black hole mass. To test this assumption, a simple and useful model can be found in Hur$\acute{\rm e}$ et al. (2011). In this work, the authors apply a linear disk model to a position-dynamical
mass diagram and re-analyze position-velocity data from H$_{2}$O maser
disks associated with active galactic nuclei. They claim that a maser
disk with nearly perfect Keplerian rotation could have disk mass
comparable to the black hole mass. This would imply that ignoring
the effects of disk self-gravity can lead to large systematic
errors in the measurement of black hole mass and the Hubble constant. 
We examine their methods and find that their large estimated disk masses of Keplerian disks
are likely the result of their 
use of projected instead of 3-dimensional position and velocity 
information. To place better constraints on the disk masses of Keplerian 
maser systems, we incorporate disk self-gravity into a 
3-dimensional Bayesian modelling program for maser disks and also
evaluate constraints based on the physical conditions for disks 
which support water maser emission. We find that there is little evidence 
that disk masses are dynamically important at the $\lesssim$1\% level 
compared to the black holes.                                                     
\end{abstract}

 
\section{Introduction} 
H$_{2}$O megamasers from circumnuclear disks (megamaser disks) in
active galaxies provide a unique way to probe active galactic nuclei.   
Megamaser disks, such as in the archetypal maser galaxy
NGC 4258 (e.g. Herrnstein et al. 1999), are typically smaller ($r\sim$0.2
pc in NGC 4258) than the gravitational sphere of influence of their
supermassive black holes (BHs; $r \sim 1$ pc in NGC 4258). This guarantees
that the gravitational potential is dominated by the central point
mass. As a result, the rotation curves of megamasers disks often
follow a nearly perfect Keplerian law, with velocity falling as the
inverse square root of radius, allowing one to easily determine
the masses of supermassive BHs ($M_{\rm BH}$) to a few percent-level
accuracy (e.g. Kuo et al. 2011; Gao et al. 2016).                    
 
In addition to BH mass measurement, the Keplerian rotation of a megamaser
disk also allows one to use the orbits of the masing gas as a standard ruler 
and determine an angular-diameter distance to a galaxy. 
This forms the basis of the Megamaser Cosmology
Project (MCP; e.g. Braatz et al. 2010) for which one attempts to make
precise determinations of the Hubble constant ($H_{0}$) by modeling
the geometrical and kinematic information of the megamaser disks (e.g. Reid et al.
2013; Kuo et al. 2013; Kuo et al. 2015; Gao et al. 2016).               
 
While many megamaser disks display nearly perfect Keplerian rotation
curves, slight deviations from Keplerian rotation have been reported 
in the literature.  For NGC 4258, Herrnstein et al. (2005) show that 
the projected rotation curve of
high-velocity masers displays a $\sim$9 km~s$^{-1}$, or 0.8\%,
flattening of the line-of-sight velocities with respect to Keplerian
motion. Careful modeling of this slight deviation can provide a 
constraint on the mass and accretion rate of the
disk (e.g. Herrnstein et al. 2005), allowing one to obtain important
physical properties for exploring AGN.                                  
 
Constraining the total mass of the maser disk ($M_{\rm D}$) can have
important implications.  An upper limit on $M_{\rm
D}$ is important for understand the warping
mechanism of a disk (e.g. Caproni et al. 2007, Martin 2008,
Ulubay-Siddiki, Gerhard, \& Arnaboldi 2009, Bregman \& Alexander 2012).  
%For the black-hole mass and Hubble constant
%$measurements, estimates of $M_{\rm
%D}$ can inform the magnitude of the systematic errors in $M_{\rm
%BH}$ and $H_{0}$ that arise from ignoring the self-gravity of the disks in the
%analysis.                                                               
Except for a few cases, such as NGC 1068 (Lodato \& Bertin 2003)
for which $M_{\rm D}$ is known to be comparable to $M_{\rm BH}$, all
previous measurements of $M_{\rm BH}$ and $H_{0}$ from megamaser disks 
with nearly perfect Keplerian rotation curves assume that $M_{\rm
D}$ is negligible in comparison with $M_{\rm BH}$. Based on this
assumption, measurements of $M_{\rm BH}$ and $H_{0}$ with at few percent level 
accuracy can be achieved when the data quality is high, and the measurement
uncertainty is dominated by measurement error. While  
$M_{\rm D} \ll M_{{\rm BH}}$ 
seems to be a good assumption for a
nearly perfect Keplerian disk, the analysis performed
by Hur$\acute{\rm e}$ et al. (2011) challenges this assumption.
 
Based on a theoretical model of accretion disks by Hur$\acute{\rm
e}$ et al. (2008), Hur$\acute{\rm e}$ et al. (2011) derive an 
expression for the dynamical mass of orbiting gas in a megamaser
disk as a function of $M_{\rm D}$, $M_{\rm BH}$, and a surface density
profile. These authors show that $M_{\rm D}$ and $M_{\rm BH}$ for
a megamaser disk system can be inferred from a \emph{position-dynamical
mass} diagram. Their most striking result from applying this technique
to seven published megamaser systems is that the claimed disk mass 
($M_{\rm D}$ = 6.2$\times$10$^{6}$ $M_{\odot}$) for UGC 3789, which 
is comparable to their estimated BH mass 
($M_{\rm BH}$ = 8.1$\times$10$^{6}$ $M_{\odot}$).
This result is quite puzzling because it contradicts the general
picture that Keplerian rotation implies the concentration of the
gravitating mass at the dynamical center of an orbit.                   
Furthermore, were the analysis of Hur$\acute{\rm e}$ et al. (2011) to be
correct, it would suggest that systematic errors in BH
mass  and the Hubble constant from applying the H$_{2}$O megamaser technique
 to Keplerian maser systems could be significantly underestimated.           
 
In order to better understand why Hur$\acute{\rm e}$'s estimated disk masses 
in Keplerian maser disks could be so large, we re-examine their analysis in Section 2. 
In Section 3, we estimate the masses of
megamaser disks more precisely by incorporating the accretion disk
model adopted by Hur$\acute{\rm e}$ et al. (2011) in the 3-dimensional
disk modelling code used by the MCP.  This allows us to evaluate
the magnitude of systematic errors in the BH mass
from ignoring disk self-gravity.
In Section 4, we discuss the implications of our results and
make conclusions.

\section{Re-examination of the Position$-$Dynamical Mass Diagram} 

A position--dynamical mass (PDM) diagram can indicate the enclosed 
gravitating mass as a function radius in an accretion disk.          
As shown in Hur$\acute{\rm e}$ et al. (2011), the dynamical mass
$\mu$ for a maser disk system can be defined as                             
\begin{equation} 
\mu = {rv^{2} \over G} \equiv \mu(\overline{\omega})~, 
\end{equation} 
where $r$ is the physical radius of a masing spot in a
circular orbit of the disk, $v$ is the orbital velocity of the masing
spot, and $\overline{\omega}$ is the normalized radius of the masing
gas. The normalized radius is defined as  $\overline{\omega}$
$=$ $r/a_{\rm out}$, where  $a_{\rm out}$ is the outer radius of
a maser disk. 
%MJR: there is a complication here, since for a disk, as opposed to
%a sherical mass distribution, V(r) is not given by the simple formula.
%So the following sentence is not strictly true and I removed it.
%From this equation, one can see that the dynamical
%mass $\mu$ is just the enclosed mass within a certain radius $r$.       
 
Hur$\acute{\rm e}$ et al. (2011) analyze the position--dynamical mass
diagrams for seven megamaser disks published before
2011 (i.e. IC 1481, UGC 3789, NGC 3393, NGC 4258, NGC 1068, NGC 4945,
and Circinus). They find that, in most of these systems, the masses of 
the accretion disks are comparable to (i.e. UGC 3789, NGC 1068), 
or substantially greater (i.e. IC 1481, NGC 3393, Circinus) than, 
the central BH masses of $\sim10^7$ $M_{\odot}$.
For the archetypal maser galaxy, NGC 4258, while their estimated disk mass 
is substantially smaller than the BH mass, the best-fit disk mass is
$\sim10^6$ $M_{\odot}$, which may be large enough to contribute to 
uncertainty in the distance determination
for this galaxy using the H$_{2}$O megamaser technique (Herrrnstein
et al. 1999, Humphyreys et al. 2013), and hence could affect
the accuracy of Hubble constant determination based on the NGC 4258
distance (e.g. Riess et al. 2016).                                      
 
%MJR: the following has been said earlier in the paper in one way or another.
%We note that a massive accretion disk is not impossible in a maser
%system such as NGC 1068, which show substantial deviation from pure
%Keplerian rotation. However, it is striking and puzzling that sub-parsec
%scale massive disks exist in systems such as UGC 3789 and NGC 4258,
%which show nearly perfect Keplerian rotation curves.                    
 
To explore why Hur$\acute{\rm e}$'s analysis allows massive disks
in UGC 3789 and NGC 4258, we wanted to reproduce their analysis, 
but could not since their position--velocity data points were 
``{\it obtained by digitalizing graphs when published ...}''. 
This is unfortunate, since the measurements of maser position and velocity for these two galaxies were available online in 
electronic form\footnote{see Table 5 in http://iopscience.iop.org/article/10.1086/512718/fulltext/ for NGC 4258}\footnote{see Table 2,3,4 in http://iopscience.iop.org/article/10.1088/0004-637X/695/1/287/meta for UGC 3789}.  
Furthermore, the authors did not publish their digitized data,
making it impossible to repeat their analysis.
So, in order to better understand the origin
of their large disk masses, we first repeated the model-fitting done 
by Hur$\acute{\rm e}$ et al. (2011) using the measurements of 
maser position and velocity from data in the original papers.  
 
\subsection{The Maser Systems for the New Analysis}
 
In our analysis using the PDM diagram for estimating the disk mass,
we will focus on the six published Keplerian maser systems (UGC 3789,
NGC 6323, NGC 6264, MRK 1419, NGC 5765b, and NGC 4258) which are the primary targets for accurate $H_{0}$ or geometric distance measurements based on the H$_{2}$O megamaser technique (Reid et al. 2013). Our main goal is to examine whether the basic assumption that disk mass is negligible compared to the BH mass in \emph{Keplerian maser disks} holds when using the megamaser technique to measure BH mass and Hubble constant. Whether or not non-Keplerian disks\footnote{In our analysis, a non-Keplerian disk is defined as a maser disk system in which the reduced $\chi^{2}$ of the rotation-curve fitting with the Keplerian rotation law is greater than 1.5. } have substantial disk masses is not the focus of this paper.
Among these six systems we examine here, UGC 3789 and NGC 4258 allow 
a direct comparison between our work and that of Hur$\acute{\rm e}$
et al. (2011).

We do not include the five maser systems, IC 1481, NGC 3393,
NGC 1068, NGC 4945, and Circinus $-$ in Hur$\acute{\rm e}$ et al.
(2011) because these systems either have complicated maser
distributions, kinematics deviating significantly from Keplerian rotation,
or the uncertainties of the maser position measurements are relatively large. 
While these factors may not prevent BH mass estimates with accuracy sufficient for understanding the
$M_{BH}-\sigma_{\star}$ relation (e.g. Ferrarese \& Merritt 2000;
Gebhardt et al. 2000; G\"{u}tekin et al. 2009; Greene et al. 2016), which is a relation in a log-log plot,  these factors make these systems non-ideal for accurate $H_{0}$ determination because
the systematic uncertainties resulting from these factors could make the $H_{0}$ determination with percentage-level accuracy difficult, especially if non-gravitation effects such as outflows are involved (e.g. Greenhill et al. 2003). Therefore, whether these systems have substantial disk masses are cosmologically less important. 

Moreover, these factors, plus the lack of centripetal acceleration measurements of maser features in these systems which constrain maser positions along the ling-of-sight in the disks, prevent us to perform the
more robust 3-dimensional modeling of the disk shown in section 3 for
determining the disk masses in these systems. We thus ignore
them in our analysis.

%However, we include the other four systems here in order to see 
%whether or not the PDM-diagram method could imply massive disks. 

%MJR: following is repetitive.
%Were these systems to have disk mass greater than $\sim$10$^{6}$
%$M_{\odot}$ (i.e. greater than $\sim$10\% of the BH masses), it would
%have important implications for the accuracy of the $H_{0}$ determination
%using the megamaser technique, which usually assumes that the disk
%mass is negligible in comparison with the BH mass. The reference
%papers from which we obtained the data are indicated in Column (8)
%in Table 1.                                                             
 
\begin{figure*}[ht] 
\begin{center} 
%\vspace*{0 cm} 
\hspace*{-1 cm} 
\includegraphics[angle=0, scale=0.5]{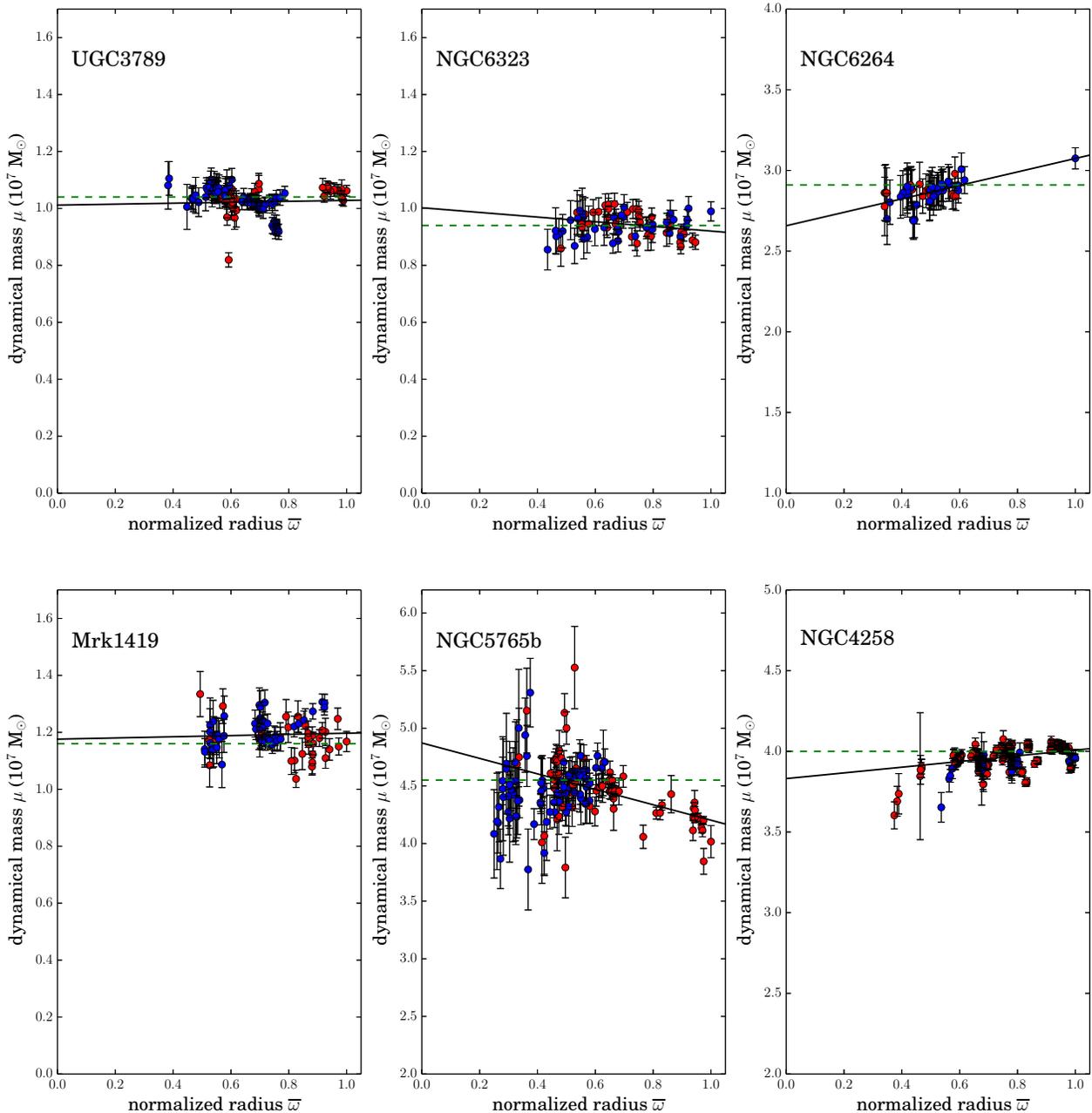} 
\vspace*{0.0 cm} 
\caption{The position$-$dynamical mass diagrams for six disk maser
systems that display nearly perfect Keplerian rotation curves. The
red and blue dots show the redshifted and blueshifted maser spots
in a disk, respectively. In each plot, the black line shows the best
linear fit to the data, and green dashed line shows black hole mass
obtained from fitting the Keplerian rotation law to the observed rotation
curve for each maser system (Kuo et al. 2011; Gao et al. 2016; Humphreys et al. 2013).
The slope of the black line represents the mass of the disk in the
context of the Hur$\acute{\rm e}$ model while its intercept with
the vertical axis of the plot indicates the black hole mass of the
system.}                                                                
\end{center} 
\end{figure*} 
 
\begin{deluxetable*}{lrrrrccl} 
%\setlength\extrarowheight{5pt}
%\tabletypesize{\scriptsize} 
\tablewidth{0 pt} 
\tablecaption{Results of linear-fitting in the position$-$ dynamical mass diagrams} 
\tablehead{ 
\colhead{Galaxy} & \colhead{Distance} & \colhead{N } & \colhead{$M_{\rm BH}$}  &  \colhead{$M_{\rm D}$}   &  \colhead{$\chi^{2}_{\nu}$} & \colhead{$\sigma_{\mu}$ } & \colhead{Reference}\\  
\colhead{Name} & \colhead{(Mpc)}     & \colhead{}   & \colhead{(10$^{7}$ $M_{\odot}$)} & \colhead{(10$^{7}$ $M_{\odot}$)} & \colhead{} & \colhead{(10$^{7}$ $M_{\odot}$)}  & \colhead{Paper}  
%& 
%  & \colhead{$i_{0}$}  & \colhead{$i_{1}$} & \colhead{$\alpha_{0}$}  &  
%\colhead{$\alpha_{1}$} & \colhead{$\alpha_{2}$} \\
%\colhead{(Mpc)} & \colhead{(10$^{7}$ $M_{\odot}$)} & \colhead{(km~s$^{-1}$)}  & \colhead{($\mu$arsec)} & \colhead{($\mu$arsec)}    & \colhead{($^\circ$)}  & \colhead{($^\circ$/mas)} &  \colhead{($^\circ$)}  &  \colhead{($^\circ$/mas)}  & \colhead{($^\circ$/mas$^{2}$)} 
}     
\startdata 
UGC 3789  & 46.4 & 113 & $1.01\pm0.04$  &        0.02$\pm$0.05   &   3.04 & 0.05  &   Reid et al. (2013)\\
NGC 6323  & 106.7 & 76  & $1.00\pm0.03$ &  $-$0.08$\pm$0.04  & 1.10    &   0.04   & Kuo et al. (2015)\\
NGC 6264  &  139.4  & 55  &  $2.66\pm0.02$  &  0.42$\pm$0.03    & 0.24  &  0.05   &  Kuo et al. (2013)\\ 
MRK 1419  &  72.2  & 76  &  $1.18\pm0.08$  &  0.02$\pm$0.09    & 2.34  &   0.06   &  Impellizzeri et al. in prep.\\ 
NGC 5765b   &  126.3     & 152  &  $4.87\pm0.06$  &  $-$0.70$\pm$0.07      & 2.19   &   0.25  &  Gao et al. (2016)\\
NGC 4258  & 7.6 & 183  &  $3.83\pm0.06$  &  0.18$\pm$0.06    & 4.56  &  0.06  &   Humphreys et al. (2013)\\ 
\enddata 
\tablecomments{  
Column (1): Name of the galaxy; column (2): galaxy distance adopted in the 
calculation of dynamical mass; column (3): number of data points for fitting; 
columns (4) and (5): the best-fit black hole and disk masses, with their 
uncertainties inflated by the square-root of the reduced $\chi^{2}$; 
column (6): the reduced $\chi^{2}$ of the fitting; column (7): the dispersion 
of the dynamical mass with respect to the linear model; column (8): references
for the positions and velocities of maser spots.
} 
\end{deluxetable*}

\subsection{The Basic Disk Model} 

The disk model derived by Hur$\acute{\rm e}$ et al. (2011) for
describing the dynamical mass distribution of an accretion disk involves
a complicated, non-linear equation that involves the BH mass ($M_{\rm
BH}$), disk mass ($M_{\rm D}$), inner/outer radius of the accretion
disk, and the surface density distribution of the disk (see equation
(6) in Hur$\acute{\rm e}$ et al. 2011).  They simplify this 
equation by assuming standard parameters of astrophysical disks, which
lead to a linear equation for the dynamical mass distribution:               
\begin{equation} 
\mu(\overline{\omega}) = M_{\rm BH} + M_{\rm D} \overline{\omega}~~. 
\end{equation} 
This shows that the observational data
plotted in a PDM diagram should display a dynamical
mass which is a linear function of the normalized radius.
By fitting a straight line to the distribution, the slope 
gives $M_{\rm D}$ and the intercept gives $M_{\rm BH}$.   
We refer to Equation (2) as the ``linear model.''

\subsection{Evaluating the Dynamical Mass }
 
Before one can fit Equation (2) to the data, one has to first compute
the dynamical mass $\mu(\overline{\omega})$ with the observed maser
positions and velocities by using Equation (1). To obtain the physical
radii of maser spots, we 
define the position of the dynamical center of the system on
the plane of the sky ($x_{0}$, $y_{0}$) by adopting the published
values from 3-dimensional disk modelling
(see the reference papers shown in Table 1).  
%MJR: isn't the following pretty obvious and hence unnecessary.
%The physical radius
%of a particular maser spot is then calculated using the projected
%distance between the maser position and the dynamical center. 
When dealing with maser position uncertainties, rather than using formal
fitting uncertainties from images, which tend
to be optimistic for high signal-to-noise data, we include more
realistic ``error floors'' adopted in the reference papers to account
for the systematic uncertainties.                
To obtain orbital velocity data, we subtract
the recession velocity of the galaxy from the observed maser velocities.
For NGC 4258 we also correct for effects of inclination of the maser disk,
which shows a $\approx8^\circ$ warping.
No disk inclination corrected are needed for the
other five systems, because those disks are within
$\approx1^\circ$ of being exactly edge-on.

\subsection{Fitting Results} 
 
Figure 1 shows the PDM diagrams for the six systems discussed here.
The red and blue dots represent the redshifted and blueshifted 
``high-velocity'' maser spots, respectively.  The solid black lines 
show the best linear fits to the data, and green dashed line shows 
black hole mass obtained from the original disk fitting, which assumed 
pure Keplerian rotation (Kuo et al. 2011; Gao et al. 2016; Humphreys et al. 2013).
These results are summarized in Table 1.  In order to make a direct
comparison with Hur$\acute{\rm e}$ et al. (2011), we first focus
on UGC 3789 and NGC 4258.                                 
 
For UGC 3789, one can see dramatic differences between the inferred
disk mass from our Figure 1 and the Hur$\acute{\rm e}$ et al. (2011) 
Figure 1.  We note that Hur$\acute{\rm e}$ used only one-third of 
the available position--velocity data.   Using the full data set,
we do not see a signifiicant slope in the PDM diagram. The best-fit 
$M_{\rm D}$ $=$ (0.02$\pm$0.05)$\times$10$^{7}$ $M_{\odot}$ is consistent 
with zero disk mass. In contrast, Hur$\acute{\rm e}$ et al. (2011) derive
a disk mass of $M_{\rm D}$ $\sim$ 0.62$\times$10$^{7}$ $M_{\odot}$,
which is comparable to their best-fit BH mass ($M_{\rm BH}$ $\sim$
0.81$\times$10$^{7}$ $M_{\odot}$). The significant difference between
these two results suggest that the large estimated disk mass for UGC 3789 
from Hur$\acute{\rm e}$ et al. (2011) is likely caused by errors from
reading data values from graphs by eye and excluding the majority of
the measured data. 
 
For NGC 4258, there is marginal evidence for a slope in the PDM diagram. 
Our best-fit $M_{\rm D}$ (0.18$\pm$0.06$\times$10$^{7}$
$M_{\odot}$) is consistent with the values from Hur$\acute{\rm
e}$ et al. ($\sim$ 0.16$\times$10$^{7}$ $M_{\odot}$), and the BH
mass differs only by 5\% between these two analysis. 
The disk mass is formally non-zero only at the 3$\sigma$ level,
and there may be some outlying data that contribute to this result.
So, before concluding that there might be a measurable disk mass
in NGC 4258, it is instructive to examine the PDM diagrams for
the other four Keplerian rotating systems.

From Figure 1, one can
see that Mrk 1419 does not have a significant slope in the PDM diagram,
and this leads to a negligible disk mass 
$M_{\rm D}$ $=$ (0.02$\pm$0.09)$\times$10$^{7}$
$M_{\odot}$.  NGC 6264 shows the most formally significant slope among the
six systems presented here, and the corresponding disk mass is 
$M_{\rm D}$ $=$ (0.42$\pm$0.03)$\times$10$^{7}$ $M_{\odot}$. On the other
hand, both NGC 6323 and NGC 5765b show negative slopes in the PDM
diagrams, implying \emph{negative} disk masses: ($M_{\rm
D}$ $=$ ($-$0.08$\pm$0.04)$\times$10$^{7}$ $M_{\odot}$ for NGC 6323
and  $M_{\rm D}$ $=$ ($-$0.70$\pm$0.07)$\times$10$^{7}$ $M_{\odot}$
for NGC 5765b). 
Note that both NGC 6323 and NGC 5765b have very well-ordered maser
disks, which do not appear to have complicated spatial structures 
(see Kuo et al. 2015; Gao et al. 2016), and their
maser kinematics can be well-fitted by Keplerian rotation. Therefore,
the negative disk masses derived from the PDM diagrams for these
two galaxies, especially for NGC 5765b, are most likely not results
from fitting maser disks with features that cannot be well-modelled
(e.g. disk thickness) or features that are suggestive more complicated
physics (e.g. outflows in Circinus; Greenhill et al. 2003).  If so,
what else can cause negative, non-physical disk mass estimates?
 
\subsection{The Origin of the Scatter in Dynamical Mass Estimates} 

In the sixth column in Table 1, one can see that four of the six
maser disk systems here have reduced $\chi^{2}_{\nu}$ values greater 
than two, which implies either that the linear model described by 
Equation (2) does not well fit the data or that the uncertainties in the 
measurements are underestimated.         
For UGC 3789, NGC 5765b, and NGC 4258, one can see
that some groups of maser spots systematically lie either above or
below the linear model.  This could be caused by ``astrophysical noise,''
indicating that not all maser spots perfectly reflect an idealized 
thin disk.   With the degree and distribution of scatter
seen in these plots, it is not unreasonable that negative disk mass
estimates can appear in certain cases.  For example, while the present 
fit for UGC 3789 shows negligible slope, were one to ignore the group of 
redshifted masers at the outer most radii, one would obtain a small
but statistically significant negative slope owing mostly to the clump
of blue shifted spots below the fitted line near a normalized radius
of 0.75.          
                              
Based on our experience modeling maser disks (e.g. Kuo et al. 2013; 
Reid et al. 2013), we argue that some of the scatter seen in the PDM
diagrams most likely originates from the differences between the
projected and 3-dimensional maser velocities and radii. When estimating
dynamical masses of maser disks using Equation (2), the Hur$\acute{\rm
e}$ et al. analysis assumes that the observed velocities and radii of 
high-velocity masers are identical to their full 3-dimensional orbital 
values.  Note that these assumptions are valid only if the high-velocity masers
lie exactly on the mid-line of the disk (i.e. the intersection
between the disk plane and plane of the sky). In reality, the high-velocity
masers in a maser disk can lie in a region that deviates from the mid-line 
of the disk by $\sim$10$-$20$^{\circ}$ (e.g. Kuo et al. 2011, Reid et al. 2013, 
Gao et al. 2016).   Because of this, the observed projected velocities of the 
outermost redshifted masers in NGC 5765b are substantially smaller than their 
true orbital velocities.  This leads to a smaller apparent dynamical mass at 
large radii, which would result in a negative disk mass based on PDM fitting.     
 
While we speculate that ignoring deviations of maser positions from the mid-line of the disk could be the primary cause of a negative disk mass, it is also conceivable that gas and radiation pressure
in the radial direction of the disk could also lead to a negative slope in a PDM diagram. This is because these pressures can provide support against gravity from the BH, making the orbital velocity smaller than the Keplerian value at a given radius and the dynamical mass will appear to be smaller than the true value. If the dynamical mass estimate that includes the pressure effect happens to decrease with radius, a negative slope in the PDM diagram will result. However, as we will show in detail in section 3.5, while radiation and gas pressure can indeed introduce a negative slope in a PDM diagram, their effect on the dynamical mass estimate is too small to explain the scatter seen in Figure 1. Therefore, we can exclude the possibility that negative disk masses can result from thermal or radiation pressure for the maser disks discussed in this paper. 
 
We conclude that while the disk model
proposed by Hur$\acute{\rm e}$ et al. (2011) can provide a simple way
to determine the BH mass and disk mass, the PDM diagram is not a
robust tool to derive accurate $M_{\rm BH}$ and $M_{\rm D}$, because
astrophyscial noise and the use of projected radii and velocities 
(instead of 3-dimensional values) can bias such dynamical mass estimates.

\section{Deriving Disk Mass from 3-dimensional  Modeling} 

A promising way to avoid bias when estimating
$M_{\rm BH}$ and $M_{\rm D}$ is to incorporate the linear model
of Hur$\acute{\rm e}$ et al. into the 3-dimensional modeling
code used by the Megamaser Cosmology Project (e.g. Humphreys et al. 2013). 
This removes projection biases and reduces systematic errors
caused by ignoring warped disk structures.                 
Before formally showing the results of $M_{\rm BH}$ and $M_{\rm D}$
measurements using the 3-dimensional model, it is helpful to first discuss 
the model fitting assuming zero disk mass. This will serve as the basis 
for comparisons, in order to see how much $M_{\rm BH}$ can change when
$M_{\rm D}$ is included in the analysis. In the following, we will
call the model in which $M_{\rm D}$ is ignored the ``fiducial'' model. 
 
\subsection{The Fiducial Model} 

The fiducial model described here was first introduced in Reid et
al. (2013) to determine accurate $H_{0}$ using the maser disk in
UGC 3789. The disk model is described by 15 global parameters including
the $H_{0}$, $M_{\rm BH}$, sky position of the dynamical center of
a maser disk, the recession velocity of the galaxy, and parameters
that described the warped structure and the eccentricity of the maser
orbits (see Reid et al. 2013 and Humphreys et al. 2013 for details).
The galaxy distance is calculated from $H_{0}$ and recessional velocity
parameters in the fitting program.                                      
 
The model assumes that the maser spots orbit around a BH with Keplerian
rotation law, and the relationship between the maser velocity $v_{\rm
orb}$ and the BH mass can be described by                               
\begin{equation} 
v_{\rm orb} =  \sqrt{{GM_{\rm BH} \over r}}~, 
\end{equation} 
where $r$ is the orbital radius of a particular maser spot. The data
that are fitted with the fiducial model are positions, velocities,
and accelerations of the maser spots in a disk. Among these data,
maser accelerations play a crucial role for constraining the spatial
location of a maser spot along the line-of-sight.                       
 
For the purpose of estimating $M_{\rm D}$ relative to $M_{\rm BH}$
in the context of the Hur$\acute{\rm e}$ et al. (2011) approach,
we held $H_{0}$ constant at a value consistent with the assumed 
angular$-$diameter distance (listed in Table 1) and recessional velocity 
for each galaxy.
The program uses a Markov chain Monte Carlo (McMC) approach,
and we adopted the median of the marginalized posteriori density functions
as best-fit parameter values, with the uncertainties 
spanning 68\% confidence intervals.
We usually generated $\ge10^{9}$ McMC trials in order to ensure
convergence of the Bayesian fitting.
The results of applying the fiducial model to fit the data are summarized 
in Table 2.

\subsection{Results from Including the Disk Model} 

Implementing the disk model of Hur$\acute{\rm e}$ et al.
in our 3-dimensional                                                    
disk modeling code is straightforward. Based on Equation (1), one can express the orbital velocity of a maser spot as a function of 
the dynamical mass and orbital radius as                    
\begin{equation} 
v_{\rm orb} =  \sqrt{{G(M_{\rm BH}+M_{\rm D}\overline{\omega}) \over r}}~.
\end{equation} 
Note that this equation is essentailly equivalent to Equation (3) if one replaces
$M_{\rm BH}$ in the equation with the dynamical mass defined
by Equation (2). Here, $M_{\rm D}$ is added in the disk model as a new global parameter. 
 
Figures 2 \& 3 show the marginalized PDFs of $M_{\rm BH}$ and $M_{\rm D}$ 
for the six maser systems considered here.  The dashed lines represent the 
PDFs of $M_{\rm BH}$ derived using the fiducial model, and the dotted lines 
show the PDFs for $M_{\rm BH}$ and $M_{\rm D}$ using the linear model, 
which allows for disk mass.  Table 2 summarizes the best-fit values
for $M_{\rm BH}$ and $M_{\rm D}$.  The reduced $\chi^{2}_{\nu}$ of the
fits now range from $\approx0.5~-~1.4$ (see column 11 in Table 2),
indicating considerable improvement over those in Table 1 where projected,
instead of 3-dimensional, velocities and radii were used. 
These figures reveal that disk masses range from about 
$10^5$ to $10^6$ $M_{\odot}$, and that most of these estimates are not
statistically significant.
In all cases, the disk masses are at least 10 times smaller than
the BH masses.   We conclude that there is no evidence for disk masses 
comparable to BH masses as suggested by Hur$\acute{\rm e}$ et al. (2011).
 
\begin{figure*}[ht] 
\begin{center} 
%\vspace*{0 cm} 
\hspace*{-1 cm} 
\includegraphics[angle=0, scale=0.5]{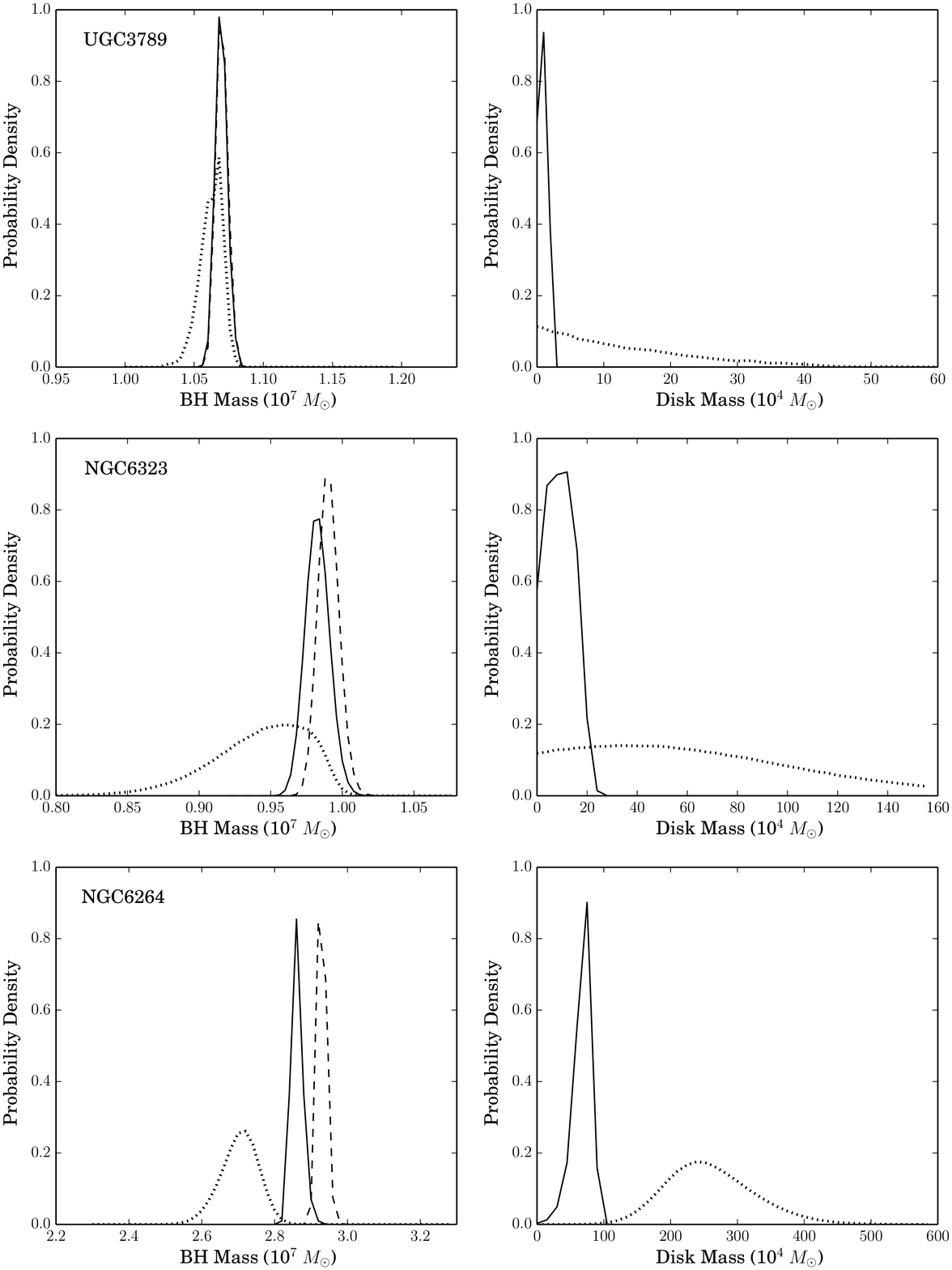} 
\vspace*{0.0 cm} 
\caption{The posteriori probability distribution functions (PDF)
of the black hole mass and disk mass from the 3-dimensional disk
modelling based on the Hur$\acute{\rm e}$ model for the Keplerian
maser disks in UGC 3789, NGC 6323, and NGC 6264. The dashed lines
shown in the left panels represent the PDFs of $M_{\rm BH}$ derived
from the fiducial model in which the disk mass is assumed to be zero.
The dotted and solid lines in the left and right panels show the
fitting results from the linear model which either including
(the solid line) or ignoring (the dotted line) the physical conditions
for maser emission in the model fitting.}                               
\end{center} 
\end{figure*} 
 
\begin{figure*}[ht] 
\begin{center} 
%\vspace*{0 cm} 
\hspace*{-1 cm} 
\includegraphics[angle=0, scale=0.5]{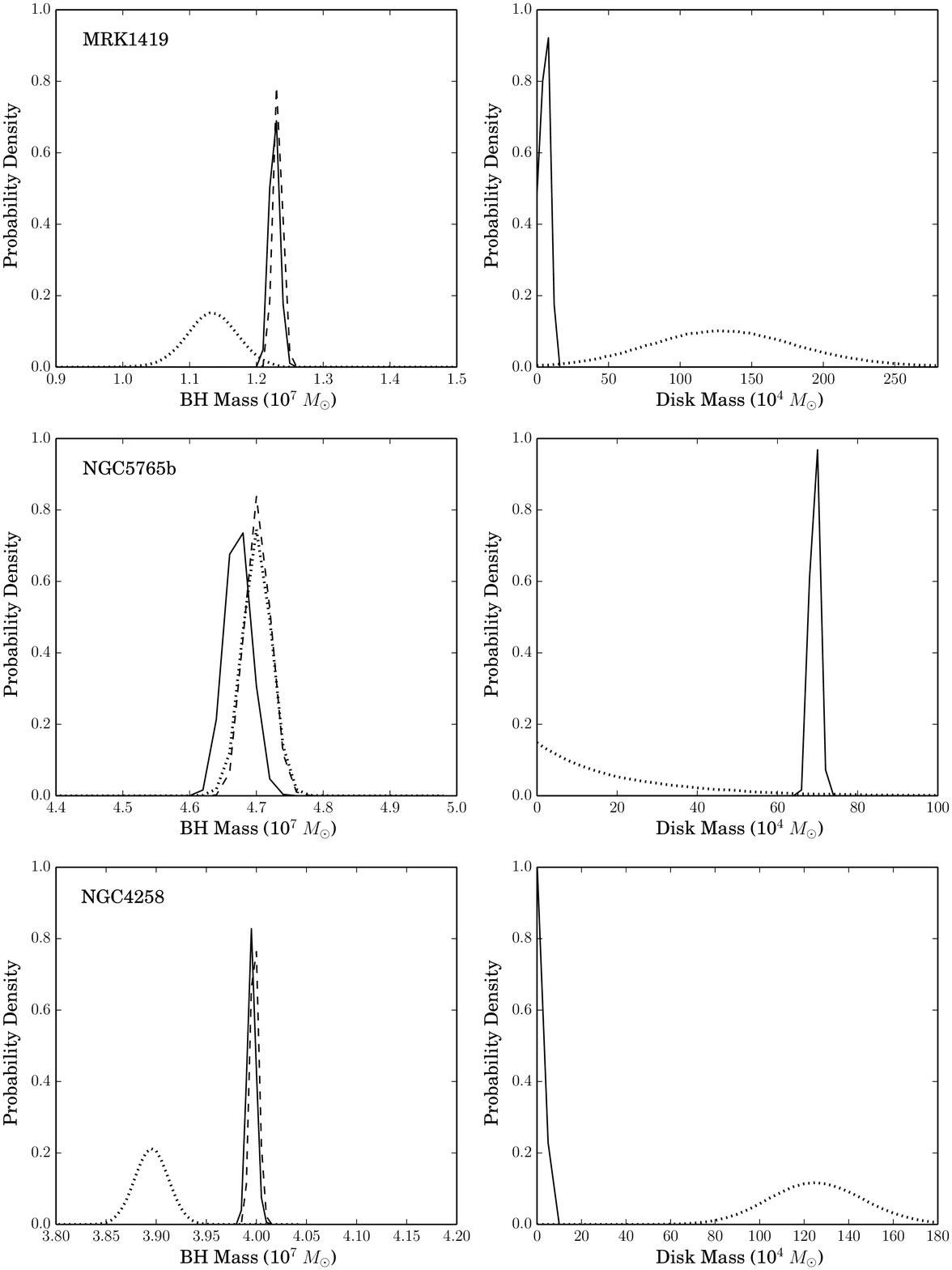} 
\vspace*{0.0 cm} 
\caption{The posteriori probability distribution functions (PDF)
of the black hole mass and disk mass from the 3-dimensional disk
modelling  based on the Hur$\acute{\rm e}$ model for the Keplerian
maser disks in MRK 1419, NGC 5765b, and NGC 4258. The dashed lines
shown in the left panels represent the PDFs of $M_{\rm BH}$ derived
from the fiducial model in which the disk mass is assumed to be zero.
The dotted and solid lines in the left and right panels show the
fitting results from the linear model which either including
(the solid line) or ignoring (the dotted line) the physical conditions
for maser emission in the model fitting.}                               
\end{center} 
\end{figure*} 
 
\subsection{Imposing Physical Conditions for Maser Emission} 

In a model that can describe H$_{2}$O maser disk satisfactorily,
an important condition that has to be satisfied is that the density of
molecular gas, n$_{\rm H_{2}}$, in the disk should be given by 
n$_{\rm H_{2}}$ $=$ $\sim10^8 - 10^{10}$ cm$^{-3}$
in order to allow sufficient amplification and avoid thermalization
of the population inversion (Herrnstein et al. 2005; Gray et al. 2016). 
Using the surface density profile 
$\Sigma$ $=$ $\Sigma(r)$ adopted by Hur$\acute{\rm e}$ et al. (2011) 
and the fitted disk mass from the previous section, one can infer 
n$_{\rm H_{2}}$ at the inner edge of a maser disk by using the equation          
\begin{equation} 
n_{\rm H_{2}} =  {\Sigma(r) \over \sqrt{2\pi} m_{\rm H_{2}}H(r)}~, 
\end{equation} 
where $m_{\rm H_{2}}$ is the mass of the molecular hydrogen, and
$H(r)$ is the scale height of the disk as a function of radius $r$.
Here, $\Sigma(r)$ $=$ $\Sigma_{\rm out}$(r/$a_{\rm out}$)$^{-1}$,
where $a_{\rm out}$ and $\Sigma_{\rm out}$ are radius and surface
density at the outer edge of a maser disk, and $\Sigma_{\rm out}$
$=$ $M_{\rm D}$/(2$\pi$$a_{\rm out}^{2}$).\footnote{Note that this expression for $\Sigma_{\rm out}$ is valid only when $a_{\rm out}$ $\gg$ $a_{\rm in}$, where $a_{\rm in}$ is the radius of the inner edge of an accretion disk and is assumed to be a few Schwarzchild radii ($R_{\rm s}$) in Hur$\acute{\rm e}$ et al. (2011). This suggests that the disk mass discussed in Hur$\acute{\rm e}$ et al. does not only account for the mass contribution from the masing region of the disk, but also mass contribution from regions well within the inner edge of a maser disk, which typically has an inner radius of $\sim$10$^{5}$ $R_{\rm s}$ (Kuo et al. 2011).    }
$H(r)$ can be estimated from
the following equation (Neufeld \& Maloney 1995):           
\begin{equation} 
H(r) =  c_{\rm s}r\sqrt{r/GM_{\rm BH}}~, 
\end{equation} 
where $c_{\rm s}$ is the sound speed of the gas in the disk; we
adopt $c_{\rm s}$ $=$ 2.15$\pm$0.15 km~s$^{-1}$, corresponding to
a gas temperature of 700$\pm$100 K in the disk modeling. We choose
100 K as the uncertainty for gas temperature because the 3$\sigma$
uncertainty of the chosen temperature distribution corresponds to
the preferred temperature range for 22 GHz H$_{2}$O maser emission of
400$-$1000 K estimated by Herrnstein et al.~(2005). Note that the three dimensional disk modeling presented here is insensitive to the 100 K temperature uncertainty. Increasing the uncertainty by a factor of 2 would only lead to a negligible change in BH mass and a change in the disk mass less than $\sim$10\%.                          
 
In columns (9) of Table 2, we show the inferred number densities at
the inner edge of the six maser disks. One can see that, except for
NGC 5765b, these values (including the ranges allowed by the uncertainties)
are in general substantially greater than the range for
maser emission.  In the case of MRK 1419 and NGC 4258, 
n$_{\rm H_{2}} > 10^{11}$ cm$^{-3}$, for which collisions would 
thermalize the level populations and quench any masing.
This suggests that the true upper bound for the disk mass must be
smaller than the values we show in the previous section.

In order to place a more physically realistic constraint on disk
mass, in our disk modeling we impose 
10$^{8} < $ n$_{\rm H_{2}} < $ 10$^{10}$ cm$^{-3}$
between the observed inner and outer edge of a maser disk. 
In Figures 2 and 3, the solid black lines shows the PDFs of $M_{\rm
BH}$ and $M_{\rm D}$ after imposing this density condition in the 
linear model, and the results of the fitting are
summarized in the third row for each galaxy in Table 2.                
This considerably reduces the disk mass estimates. 
In this new fitting, most disk masses are
all below 10$^{5}$ $M_{\odot}$.   The fractional changes in BH mass
with respect to the fiducial model are 0.0\%, 0.8\%, 2.3\%, 0.4\%,
0.6\%, and 0.05\% for UGC 3789, NGC 6323, NGC 6264, MRK 1419, NGC
5765b, and NGC 4258, respectively.  Since these uncertainties are
in general smaller than the measurement errors for these galaxies,
this implies that when a more realistic model for maser disks is
considered, the impact of the self-gravity of the disk in the modeling
is negligible for these systems.

 \begin{deluxetable*}{lrccccccccc} 
%\setlength\extrarowheight{5pt}
%\tabletypesize{\scriptsize} 
\tablewidth{0 pt} 
\tablecaption{Results of Bayesian Disk Fitting} 
\tablehead{ 
\colhead{Galaxy} & \colhead{Model} & \colhead{Masing} & \colhead{$M_{\rm BH}$}  & \colhead{$\dot{M}$ } & \colhead{$M_{\rm D}$}  &  \colhead{$R_{in} $}   &  \colhead{$R_{out} $} & \colhead{$n_{in}$ } & \colhead{$n_{out}$} & \colhead{$\chi^{2}_{\nu}$} \\  
\colhead{Name}      & \colhead{}   & \colhead{Condition}  & \colhead{(10$^{7}$ $M_{\odot}$)} & \colhead{(10$^{-4}$ $M_{\odot}$~yr$^{-1}$)} & \colhead{(10$^{7}$ $M_{\odot}$)} & \colhead{(pc)}& \colhead{(pc)} & \colhead{(10$^{9}$ cm$^{-3}$)}  & \colhead{(10$^{9}$ cm$^{-3}$)}  &  \colhead{} 
%& 
%  & \colhead{$i_{0}$}  & \colhead{$i_{1}$} & \colhead{$\alpha_{0}$}  &  
%\colhead{$\alpha_{1}$} & \colhead{$\alpha_{2}$} \\
%\colhead{(Mpc)} & \colhead{(10$^{7}$ $M_{\odot}$)} & \colhead{(km~s$^{-1}$)}  & \colhead{($\mu$arsec)} & \colhead{($\mu$arsec)}    & \colhead{($^\circ$)}  & \colhead{($^\circ$/mas)} &  \colhead{($^\circ$)}  &  \colhead{($^\circ$/mas)}  & \colhead{($^\circ$/mas$^{2}$)} 
}     
\startdata 
UGC 3789  & Fiducial  & ---  &  1.071$^{+0.004}_{-0.004}$ &  ---  & ---  & --- &  --- & --- & ---  &  0.642  \\
UGC 3789 & Linear & No   & 1.065$^{+0.007}_{-0.009}$ & ---   &  0.010$^{+0.014}_{-0.008}$ & 0.075  & 0.203  & 40$^{+60}_{-33}$  &  3.3$^{+4.8}_{-2.7}$ &   0.626 \\
UGC 3789  & Linear & Yes  & 1.071$^{+0.004}_{-0.004}$ & ---   & 0.0013$^{+0.0007}_{-0.0007}$  &  0.075 & 0.203 & 6$^{+3}_{-3}$  &  0.5$^{+0.3}_{-0.3}$ & 0.697   \\
UGC 3789 & Herrnstein & No  &  1.068$^{+0.005}_{-0.006}$ &  71$^{+100}_{-53}$   &  0.007$^{+0.001}_{-0.005}$ & 0.075  & 0.203  & 65$^{+90}_{-48}$  &  3.2$^{+4.3}_{-2.4}$ &   0.600 \\
UGC 3789  &  Herrnstein & Yes  & 1.071$^{+0.004}_{-0.004}$  &  7$^{+4}_{-3}$  & 0.0007$^{+0.0003}_{-0.0003}$  & 0.075  & 0.203 & 6$^{+3}_{-3}$  & 0.3$^{+0.1}_{-0.1}$  & 0.595   \\
\hline
NGC 6323  &  Fiducial & --- & 0.992$^{+0.007}_{-0.007}$  &  ---  & ---   & --- & --- & ---  & --- &  0.514 \\
NGC 6323  & Linear & No    & 0.949$^{+0.029}_{-0.039}$  &  --- &  0.061$^{+0.058}_{-0.041}$ & 0.144 & 0.309 & 31$^{+29}_{-21}$  &  4.5$^{+3.9}_{-3.0}$ & 0.513  \\
NGC 6323  &  Linear & Yes    & 0.984$^{+0.008}_{-0.008}$  &  ---  &  0.011$^{+0.06}_{-0.06}$ & 0.145 & 0.306 & 6$^{+3}_{-3}$  & 0.8$^{+0.5}_{-0.5}$  & 0.502   \\
NGC 6323  &  Herrnstein & No   &  0.976$^{+0.013}_{-0.015}$ & 346$^{+326}_{-228}$   &  0.034$^{+0.032}_{-0.027}$ & 0.144  & 0.309 & 41$^{+39}_{-27}$  & 4.1$^{+3.5}_{-2.7}$ &   0.491 \\
NGC 6323  &  Herrnstein & Yes   & 0.990$^{+0.007}_{-0.007}$  & 48$^{+31}_{-27}$   & 0.005$^{+0.003}_{-0.003}$  & 0.145  & 0.306  & 6$^{+3}_{-3}$ &  0.6$^{+0.3}_{-0.3}$ & 0.470   \\
\hline
NGC 6264  & Fiducial  & --- & 2.939$^{+0.012}_{-0.010}$  & ---  & ---  &  --- & --- & --- & --- & 0.778   \\
NGC 6264  & Linear & No  & 2.716$^{+0.047}_{-0.054}$  & ---  & 0.262$^{+0.071}_{-0.059}$  &  0.265 & 0.483  &  31$^{+9}_{-7}$ &  6.9$^{+1.8}_{-1.5}$ & 0.671   \\
NGC 6264  &  Linear & Yes & 2.870$^{+0.016}_{-0.013}$  & ---  & 0.077$^{+0.010}_{-0.015}$ & 0.268  & 0.481  & 9$^{+1}_{-2}$  &  2.1$^{+0.2}_{-0.4}$ & 0.724  \\
NGC 6264  &  Herrnstein & No   &  2.853$^{+0.018}_{-0.018}$ & 749$^{+212}_{-178}$   &  0.128$^{+0.031}_{-0.227}$ & 0.264  &  0.483 & 41$^{+10}_{-9}$  & 6.7$^{+1.6}_{-1.4}$  & 0.664   \\
NGC 6264  &  Herrnstein & Yes    & 2.916$^{+0.012}_{-0.010}$  & 184$^{+51}_{-51}$   & 0.030$^{+0.004}_{-0.007}$ & 0.268  & 0.481  & 10$^{+1}_{-1}$  & 1.6$^{+0.2}_{-0.4}$ & 0.771   \\
\hline
Mrk 1419  & Fiducial  & ---   &  1.237$^{+0.007}_{-0.006}$ & ---   & ---  & ---  & ---  &  --- &  --- &  1.375 \\
Mrk 1419  & Linear & No   & 1.139$^{+0.037}_{-0.037}$  & ---  & 0.133$^{+0.054}_{-0.051}$  & 0.121 & 0.310  & 111$^{+48}_{-44}$  & 10.6$^{+4.0}_{-3.9}$  & 1.373   \\
Mrk 1419  & Linear  & Yes  & 1.232$^{+0.007}_{-0.007}$  &  --- &  0.008$^{+0.004}_{-0.004}$ &  0.126 & 0.307  & 6$^{+3}_{-3}$  & 0.6$^{+0.3}_{-0.4}$ &  1.362 \\
Mrk 1419  &  Herrnstein & No   &  1.187$^{+0.020}_{-0.020}$ & 637$^{+285}_{-252}$   & 0.082$^{+0.034}_{-0.032}$  & 0.121  & 0.310  & 152$^{+69}_{-62}$  & 9.0$^{+3.4}_{-3.3}$  &  1.064 \\
Mrk 1419  & Herrnstein & Yes   & 1.235$^{+0.007}_{-0.006}$  & 36$^{+18}_{-17}$   & 0.004$^{+0.002}_{-0.002}$  & 0.126  & 0.307 &  8$^{+3}_{-4}$ &  0.5$^{+0.2}_{-0.3}$ & 1.064   \\
\hline
NGC 5765b  &  Fiducial & ---   &  4.712$^{+0.018}_{-0.018}$ & ---   & ---  & --- & ---  & ---  & --- &  0.894 \\
NGC 5765b  &  Linear & No  & 4.711$^{+0.021}_{-0.021}$  & ---   & 0.013$^{+0.024}_{-0.011}$ &  0.302 &  1.252 & 0.5$^{+0.9}_{-0.4}$ &  0.02$^{+0.03}_{-0.01}$ & 0.770   \\
NGC 5765b  &  Linear & Yes  &  4.682$^{+0.019}_{-0.019}$ & ---  &  0.070$^{+0.001}_{-0.001}$ &  0.326 & 1.236 & 2.9$^{+0.2}_{-0.1}$  &  0.1$^{+0.004}_{-0.002}$ & 0.997  \\
NGC 5765b  & Herrnstein & No &  4.713$^{+0.020}_{-0.020}$ &  30$^{+50}_{-23}$  & 0.020$^{+0.033}_{-0.015}$  & 0.321 & 1.233 &  1.5$^{+2.6}_{-1.2}$ &  0.03$^{+0.04}_{-0.02}$ & 0.737  \\
NGC 5765b  &  Herrnstein & Yes   &  4.663$^{+0.019}_{-0.018}$ &  107$^{+32}_{-24}$  & 0.068$^{+0.017}_{-0.009}$  & 0.332 & 1.204 &  5.0$^{+1.3}_{-0.7}$ & 0.1$^{+0.03}_{-0.01}$  & 0.989  \\
\hline
NGC 4258  &  Fiducial & ---  &  4.000$^{+0.004}_{-0.004}$ & ---  & ---  & --- &---  &  --- & ---  & 0.555  \\
NGC 4258  &  Linear & No   & 3.898$^{+0.017}_{-0.017}$  & ---  &  0.128$^{+0.022}_{-0.021}$& 0.114 & 0.301  & 239$^{+45}_{-41}$  & 21.0$^{+3.9}_{-3.6}$  & 0.490  \\
NGC 4258  &  Linear & Yes  &  3.998$^{+0.004}_{-0.004}$ & ---   &  0.004$^{+0.001}_{-0.002}$ &  0.115 & 0.299  &  7$^{+2}_{-3}$ & 0.7$^{+0.2}_{-0.3}$  & 0.550   \\
NGC 4258  &  Herrnstein & No &  3.936$^{+0.010}_{-0.010}$ &  380$^{+84}_{-74}$  &  0.089$^{+0.014}_{-0.013}$ & 0.113  &  0.300 &  362$^{+64}_{-59}$ &  19.5$^{+3.3}_{-3.1}$ &  0.498 \\
NGC 4258  &Herrnstein & Yes   &  4.000$^{+0.004}_{-0.004}$ & 8$^{+3}_{-3}$   &  0.002$^{+0.001}_{-0.001}$ & 0.114  & 0.299  & 7$^{+2}_{-3}$  & 0.4$^{+0.1}_{-0.2}$ & 0.566   \\
\enddata 
\tablecomments{  
Column (1): Name of the galaxy; column (2): the accretion disk model used for 
Bayesian fitting; column (3): whether the physical conditions for 22~GHz water 
maser emission were imposed in the fitting; column (4): the best-fit black 
hole mass; column (5): the best-fit mass accretion rate, normalized by 
$\alpha$, the Shakura-Sunyaev viscosity parameter (Shakura \& Sunyaev 1973). Here, we set $\alpha = 1$; 
column (6): the best-fit disk mass; columns (7) and (8): the inner and outer
radii of the maser disk; column (9): columns (10) and (11): 
the number densities of molecular hydrogen at the inner and outer edges of 
the maser disk; column (12) the reduced $\chi^{2}$ of the fit. 
Distances adopted in the Bayesian fitting for UGC 3789, NGC 6323, NGC 6264, 
MRK 1419, NGC 5765b, and NGC 4258 were 46.4, 106.7, 139.4, 72.2, 126.3, 7.6 
Mpc, respectively.
} 
\end{deluxetable*} 

\subsection{Comparison with the Herrnstein Disk Model} 

In the previous section, we saw that the linear model gave negligible
disk masses when imposing the density condition
for maser emission in the disk fitting. It would be interesting to
see whether different accretion disk models give similar results. 
 
\begin{figure*}[ht] 
\begin{center} 
%\vspace*{0 cm} 
\hspace*{-1 cm} 
\includegraphics[angle=0, scale=0.5]{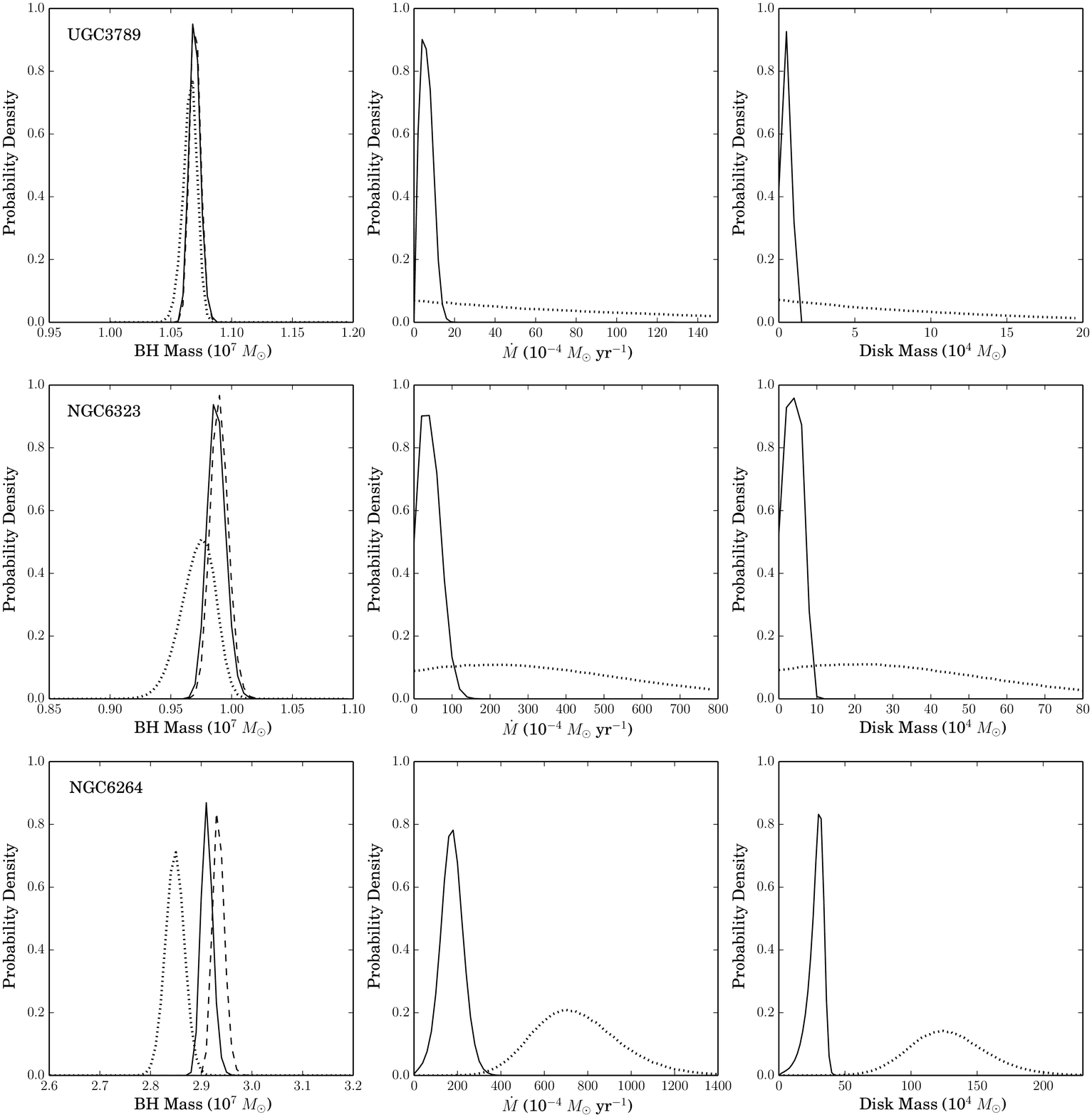} 
\vspace*{0.0 cm} 
\caption{The posteriori probability distribution functions (PDF)
of the black hole masses, mass accretion rates, and disk masses from the
3-dimensional disk modelling based on the Herrnstein model for the
Keplerian maser disks in UGC 3789, NGC 6323, and NGC 6264. The dashed
lines shown in the left panels represent the PDFs of $M_{\rm BH}$
derived from the fiducial model in which the disk mass is assumed
to be zero. The dotted and solid lines in the left and right panels
show the fitting results from the Herrnstein model which either including
(the solid line) or ignoring (the dotted line) the physical conditions
for maser emission in the model fitting.}                               
\end{center} 
\end{figure*} 
 
\begin{figure*}[ht] 
\begin{center} 
%\vspace*{0 cm} 
\hspace*{-1 cm} 
\includegraphics[angle=0, scale=0.5]{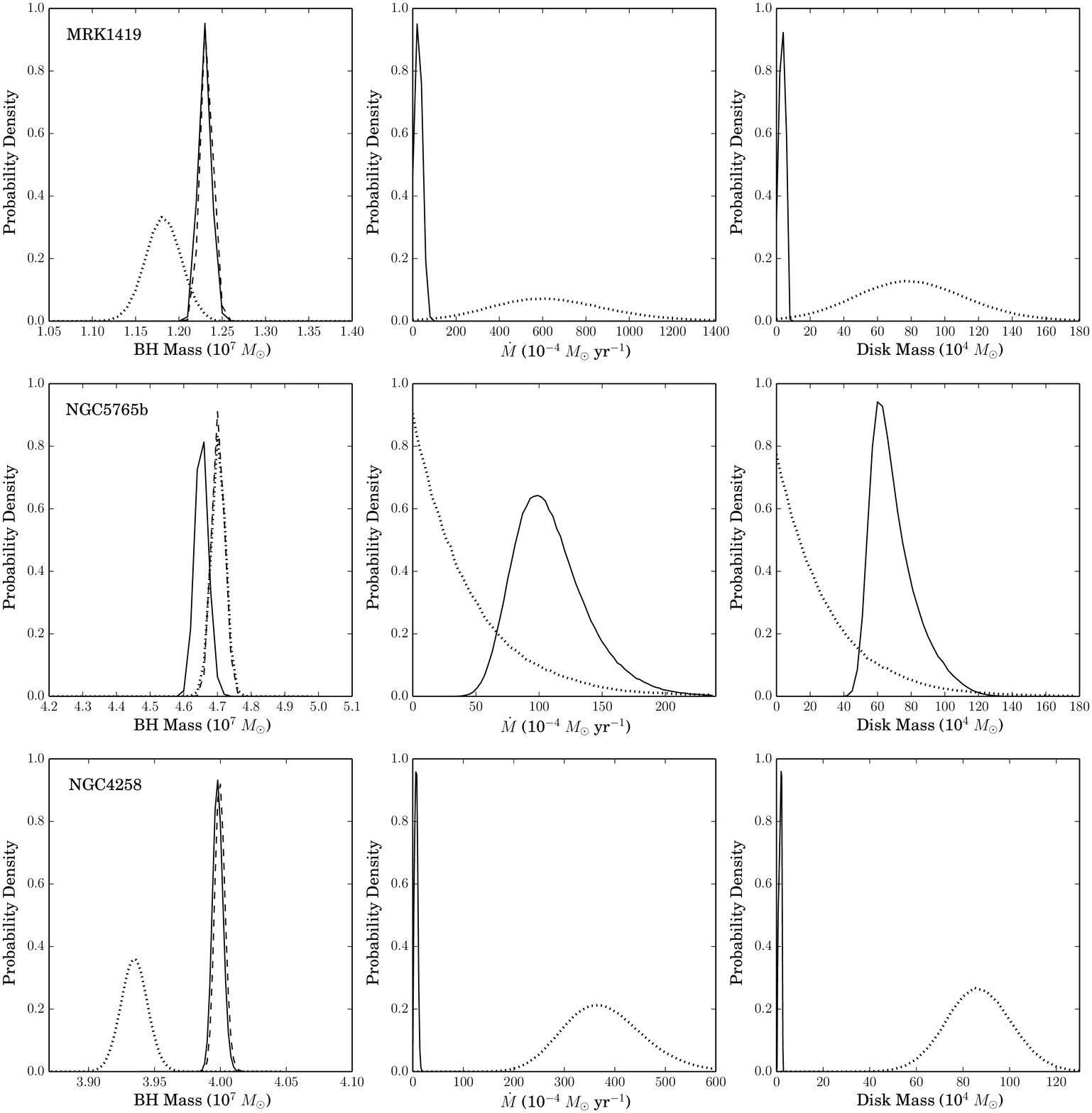} 
\vspace*{0.0 cm} 
\caption{The posteriori probability distribution functions (PDF)
of the black hole masses, mass accretion rates, and disk masses from the
3-dimensional disk modelling based on the Herrnstein model for the
Keplerian maser disks in UGC 3789, NGC 6323, and NGC 6264. The dashed
lines shown in the left panels represent the PDFs of $M_{\rm BH}$
derived from the fiducial model in which the disk mass is assumed
to be zero. The dotted and solid lines in the left and right panels
show the PDFs for $M_{\rm BH}$ and $M_{\rm D}$ from the Hur$\acute{\rm
e}$ model. The dotted and solid lines in the left and right panels
show the fitting results from the Herrnstein model which either including
(the solid line) or ignoring (the dotted line) the physical conditions
for maser emission in the model fitting.}                               
\end{center} 
\end{figure*} 
 
The disk model we use here for comparison was used by
Herrsntein et al. (2005) to describe the maser disk in NGC 4258.
In this model, which assumes steady-state accretion, the mid-plane
density of the accretion disk $\rho_{\rm mid}$ follows the equation:    
\begin{equation} 
\rho_{\rm mid}(r) = {GM_{\rm BH}\dot{M} \over 3\pi(2\pi)^{1/2}\alpha
r^{3}c_{s}^{3}}~,                                                       
\end{equation} 
where $\dot{M}$ is the mass accretion rate, $\alpha$ is the Shakura-Sunyaev
viscosity parameter (Shakura \& Sunyaev 1973), $r$ is radius, 
and $c_{s}$ is the sound speed. Given the disk scale
height described by $H = c_{s}r(r/GM_{\rm BH})^{1/2}$,                  
one can show that the surface density profile $\Sigma$ is proportional
to $r^{-3/2}$, which differs from that adopted for
the linear model (i.e. $\Sigma \propto r^{-1}$) in Hur$\acute{\rm
e}$ et al. (2011).                                                      
 
Using Equation (5), Herrnstein et al. (2005) derive the mass distribution 
$M_{\rm D}(r)$ in the disk :                  
\begin{equation} 
M_{\rm D}(r) = 8.3\times10^{4}{\dot{M} \over \alpha}{M_{\rm BH}^{1/2}(r^{1/2}-r_{\rm
min}^{1/2}) \over c_{s}^{2}}~M_{\odot}~,                                
\end{equation} 
where $r_{\rm min}$ is the physical radius of the innermost maser
spot in the disk.                                                       
 
To implement disk mass in the 3-dimensional Bayesian modeling code,
we follow the approach from Herrnstein et al. (2005) and replace
$M_{\rm BH}$ in Equation (3) by $M_{\rm BH}$ $+$ $M_{\rm D}(r)$ :       
\begin{equation} 
v_{\rm orb} =  \sqrt{{G[M_{\rm BH}+M_{\rm D}(r)] \over r}}~. 
\end{equation} 
Here, $c_{s}$ in Equation (8) is added in the model with a Gaussian
prior of 2.15$\pm$0.15 km~s$^{-1}$.                                              
We also require n$_{\rm H_{2}}$ to be within 10$^{8}$$-$10$^{10}$
cm$^{-3}$ in the disk modelling.                                        
 
Figures 4 \& 5 shows the PDFs for $M_{\rm BH}$, $\dot{M}$, and $M_{\rm
D}$ for the six Keplerian maser disks and the fitting results are
summarized in the fourth and fifth rows for each galaxy in Table
2. Similar to Figures 2 \& 3, dashed lines in the figures show the
BH mass measurements based on the fiducial model. The solid and dotted
lines represent measurements from the Herrnstein model with and without
including constraints on gas density in the model fitting, respectively.
These figures show that in all cases, the disk mass is $\lesssim$1\% of 
the BH mass, and the fractional changes in
BH mass estimates also are all $\lesssim$1\%. 
These results are consistent with $M_{\rm D}$ estimtes in Section 3.3 using
the linear model with a density constraint.
 
Note that while the accretion disk model we adopt here is valid only
for an accretion disk in steady-state, which may not be a valid
assumption for a maser disk in reality because the timescale for
such a disk to achieve steady state can be as long as a few $\times$10$^{9}$
years (e.g.  Gammie, Narayan, and Blanford 1999), the consistent
$M_{\rm BH}$ and $M_{\rm D}$ derived from the linear model
and the Herrnstein model suggests that our model fitting is not 
highly sensitive to the actual density distribution of the disk. 
 
\subsection{The Effect of Gas and Radiation Pressure on Maser Disks}
One important assumption that has been made in the disk modeling discussed above is that the dynamical effect of
gas and radiation pressure is negligible in comparison with gravity from the BH. This assumption can be tested by including the pressure terms
in Equation (3) (Haworth et al. 2018) :    
\begin{equation} 
v_{\rm orb} =  \sqrt{{GM_{\rm BH} \over r} - {f_{\rm rad}r \over \rho} + {r \over \rho}{dP_{\rm gas} \over dr} }~,
\end{equation} 
where $f_{\rm rad}$ is the radiation pressure force per unit volume, $\rho$ is the local volume density, and $P_{\rm gas}$ is the thermal pressure of the gas in the maser disk.
Assuming that the standard picture of maser excitation in an accretion disk (Neufeld, Maloney, Conger 1994; Neufeld \& Maloney 1995) is correct,  the mid-plane gas in a warped maser disk is directly illuminated by the AGN radiation obliquely which heat the gas to sufficient temperature for maser excitation. Assuming that the AGN radiation is isotropic and totally absorbed by the maser disk, we can thus re-write the second term in 
equation (10) as 
\begin{equation}
{f_{\rm rad}r \over \rho} = {L_{\rm bol} \over 4\pi crm_{H_{2}}n_{\rm H_{2}}(r)H(r)}sin(\alpha(r))~,
\end{equation}
where $L_{\rm bol}$ is the bolometric luminosity of the AGN, c is the speed of light, and $\alpha (r)$ is the angle between the warped plane of the maser disk at radius r and the incident light ray that hits the disk plane. Since all maser disks discussed in this paper are only slightly warped, the angle $\alpha$ is typically only a few degrees (e.g. Reid et al. 2013; Gao et al. 2016).
 
Assuming that the density distribution follows Equation (5) and that the temperature does not vary in the masing region of the disk, one can express the third term of Equation (10) as
\begin{equation}
{r \over \rho}{dP_{\rm gas} \over dr} = -{5 \over 2}{P \over \rho} = -{5 \over 2}{kT \over m_{H_{2}}}~.
\end{equation}
To see the effect of these two pressure terms on the dynamical mass measurement more clearly, it is helpful to re-express Equation (10) in the form of dynamical mass :
\begin{equation}
\mu \equiv {rv_{\rm orb}^{2} \over G} = M_{\rm BH} - \Delta M_{\rm rad} - \Delta M_{\rm gas}~,
\end{equation}
where $\Delta M_{\rm rad}$ is defined as 
\begin{equation}
\Delta M_{\rm rad} = {L_{\rm bol} \over 4\pi cGm_{H_{2}}n_{\rm H_{2}}(r)H(r)}sin(\alpha(r)) ~,
\end{equation}
and $\Delta M_{\rm gas}$ is defined as
\begin{equation}
\Delta M_{\rm gas} = {5 \over 2}{kTr \over Gm_{H_{2}}}~.
 \end{equation}
Assuming that $H(r)$ follows Equation (6), the BH mass is $10^{7}$ $M_{\odot}$, the gas temperature T is 700 K and $\alpha$ is $3^{\circ}$, the inner radius of the maser disk is 0.2 pc, and the number density n$_{\rm H_{2}}$(r) at the inner radius is 10$^{10}$ cm$^{-3}$, we can re-express Equation (14) \& (15) as
\begin{equation}
\Delta M_{\rm rad} = 1095 \Big({L_{\rm bol} \over 10^{44} {\rm erg~s^{-1}}}\Big)\Big({r \over 0.2 {\rm pc}}\Big) ~M_{\odot}~,
\end{equation}
\begin{equation}
\Delta M_{\rm gas} = 336 \Big({T \over 700 {\rm K}}\Big)\Big({r \over 0.2 {\rm pc}}\Big) ~M_{\odot}~.
\end{equation}
These equations show that the gas and radiation pressure can indeed introduce a negative slope in a PDM diagram, leading to negative disk mass. However, given that the bolometric luminosities of AGN in the six maser galaxies range from $\sim$1$\times$10$^{42}$ $-$ 6$\times$10$^{44}$ erg~s$^{-1}$ (Herrnstein et al. 2005; Greene et al. 2010; Gao et al. 2017), and the sizes of the maser disks are typically less than 1 pc, one can easily show that 
the pressure effects can only affect the dynamical mass estimates at the levels between 2 to 5 orders of magnitude lower than the black masses and cannot explain the scatter in the PDM diagram seen in Figure 1. From the above discussion, we can conclude that the thermal and radiation pressure are dynamically unimportant in comparison with the gravity of the BHs and can thus be ignored. 
 
\section{Discussion and Summary}

The H$_{2}$O megamaser technique provides a unique way to determine
the Hubble constant and BH mass with high accuracy, and systematic
uncertainties in these parameters are thought to be small.  The effectiveness
of this method comes from the observation that the physics of the maser 
disk is relatively simple, clean, and well understood. There are only a 
few assumptions in the disk modeling:

\begin{itemize} 
\item[$\bullet$] The masing gas follow circular orbits around the
central BH of the disk.                                                 
 
\item[$\bullet$] The dynamics of the masing gas is dominated by the
gravity of the BH and the effect of the self$-$gravity of the maser
disk can be ignored. As a result, the gas follow Keplerian rotation
\emph{exactly} with negligible degree of deviation.                     
 
\item[$\bullet$] The maser kinematics are minimally affected by 
non-gravitational forces such as radiation pressure (e.g. Maloney, Begelman,
\& Pringle 1996) or shocks from spiral density waves in the disk
(e.g. Maoz \& McKee 1998).                                              
 
\end{itemize} 

The first assumption can be directly tested by allowing orbital eccentricity
($e$) to be a free parameter in the 3-dimensional modelling program. 
Based on results to date, there is no evidence for significant eccentricities 
(i.e. $e$ $>$ 0.1), which could introduce significant systematics in 
parameter estimation (e.g. Reid et al. 2013; Humphreys et al. 2013).
Therefore, this assumption has been well tested.
 
The existence of spiral density waves in maser disks was suggested 
by Maoz \& McKee (1998), owing to some potential periodicities observed
in the spatial distribution of high-velocity masers toward NGC 4258.
Pesce et al. (2015) evaluated several sources and concluded that there 
is little evidence for spiral structure for most published Keplerian 
maser disks. This suggests
that effects of the putative spiral density waves in maser disks
on $H_{0}$ determination is most likely to be negligible.              
 
The only assumptions that have not been previously examined 
in some detail are the possible effects of radiation/gas pressure and the 
impact of self-gravity of maser disks on disk parameter estimation.
 
Ignoring self-gravity was expected to introduce negligible systematic
errors ($\lesssim$1\%) in the modeling, as the masers often displayed
near-perfect Keplerian motions.  However, the work done by 
Hur$\acute{\rm e}$ et al. (2011) challenged this point of view. 
These authors suggested that disk masses in the maser galaxies NGC 4258
and UGC 3789 are a few times 10$^{6}$ $M_{\odot}$, with the disk
mass comparable to the BH mass in UGC 3789.         
 
To understand why a Keplerian disk can have substantial disk mass,
we re-examinee the PDM diagram proposed by Hur$\acute{\rm e}$ et al.
(2011). We show that while the linear disk model proposed by Hur$\acute{\rm e}$ et al. (2011) provides 
an easy and useful method to describe the self-gravity effect of the maser disk, the PDM diagram may not be an appropriate tool to infer 
BH mass and disk mass, because the maser data are projected on the sky and one must
allow for deviations from the simplest geometry by modeling in 
3-dimensions.   Ignoring these issues can result in uncertain disk
masses that sometimes yield nonphysical (negative) disk masses.

 Despite the possibility that negative disk masses inferred from PDM diagrams could result from gas and radiation pressure,
our discussion in section 3.5 demonstrates that the pressure effects are dynamically unimportant compared to the gravity of the BHs, suggesting that negative slopes in PDM diagrams cannot be caused by the pressure effects for the maser systems discussed in this paper.

Our analysis, described in Section 3.2, show that a full 3-dimensional 
modeling, which allows for disk mass, yields small values. 
When we include physical constraints on density in the disk, in order
to allow water maser emission, the disk mass estimates drop to very
low levels, hence verifying an important assumption made in the megamaser technique for BH mass and $H_{0}$ determination based on Keplerian maser disks.

\acknowledgements We thank the anonymous referee for providing valuable suggestions for improving this paper. 
This publication is supported by Ministry of Science
and Technology, R.O.C.                                                  
under the project 104-2112-M-110-014-MY3. This 
research has made use of NASA's Astrophysics Data System Bibliographic 
Services, and the NASA/IPAC Extragalactic Database (NED) which is 
operated by the Jet Propulsion Laboratory, California Institute of 
Technology, under contract with the National Aeronautics and Space 
Administration.

\end{document}